\documentclass[journal]{IEEEtran}
\usepackage{epsfig}
\usepackage{amssymb,amsmath,amsfonts,amsthm}
\usepackage{subfigure}
\usepackage{graphicx}
\usepackage{times}
\usepackage{cite}
\usepackage{enumerate}

\usepackage{algorithm}
\usepackage{algorithmic}
\usepackage{multirow}

\usepackage{braket}
\usepackage{multicol}

\newtheorem{theorem}{Theorem}[section]

\newtheorem{definition}{Definition}[section]

\newtheorem{remark}{Remark}[section]

\begin{document}

\title{Wireless Network-Level Partial Relay Cooperation: A Stable Throughput Analysis}

\author{\IEEEauthorblockN{Nikolaos Pappas, Jeongho Jeon, Di Yuan, Apostolos Traganitis, Anthony Ephremides\thanks{N. Pappas and Di Yuan are with the Department of Science and Technology, Link\"{o}ping University, Norrk\"{o}ping SE-60174, Sweden (e-mail: nikolaos.pappas@liu.se).}
\thanks{J. Jeon is with Intel Corporation, Santa Clara, CA 95054 USA (email: jeongho.jeon@gmail.com).}
\thanks{A. Ephremides is with the Department of Electrical and Computer Engineering and Institute for Systems Research, University of Maryland, College Park, MD 20742. A. Ephremides is also with the Department of Science and Technology, Link\"{o}ping University, Norrk\"{o}ping SE-60174, Sweden, (e-mail: etony@umd.edu).}
\thanks{A. Traganitis is with the Computer Science Department, University of Crete, Greece and Institute of Computer Science, Foundation for Research and Technology - Hellas (FORTH) (e-mail: tragani@ics.forth.gr).}
}
\thanks{This work was presented in part in the IEEE International Symposium on Information Theory (ISIT) 2012 \cite{Partial-ISIT}.}
}

\maketitle

\begin{abstract}
In this work, we study the benefit of partial relay cooperation. We consider a two-node system consisting of one source and one relay node transmitting information to a common destination. The source and the relay have external traffic and in addition, the relay is equipped with a flow controller to regulate the incoming traffic from the source node. The cooperation is performed at the network level. A collision channel with erasures is considered. We provide an exact characterization of the stability region of the system and we also prove that the system with partial cooperation is always better or at least equal to the system without the flow controller.
\end{abstract}


\section{Introduction}
\label{sec:intro}

Cooperative communication helps overcome fading and attenuation in wireless networks. Its main purpose is to increase the communication rates across the network and to increase the reliability of time-varying links \cite{CoopNOW, ZlatanovCOMMAG2014}. It is known that wireless communication from a source to a destination can benefit from the cooperation of nodes that overhear the transmission. The classical single relay channel \cite{b:Muelen} exemplifies this situation. Further work on the relay channel in~\cite{b:CoverGamal} and~\cite{b:Gamal} has enabled substantial performance improvements. 

However, there is evidence that additional gains can be achieved with ``network-layer" cooperation (or packet-level cooperation), that is plain relaying without any physical layer considerations~\cite{b:Sadek} and~\cite{b:Rong1}. In this work, we focus on this type of cooperation. The work in~\cite{b:Rong3} investigated the network-level cooperation in a network consisting of a source and a relay by considering the cases of full or no cooperation at the relay.

A key difference between physical-layer and network-layer cooperation is that, for the latter, the objective rate function that is maximized is the so-called stable throughput region which captures the bursty nature of traffic from the source. In \cite{b:Rong3}, it was shown that the stability region of full cooperation under random-access does not always strictly contain the non-cooperative stability region.

Among distributed communication protocols, we are interested in ALOHA, a simple random access scheme in which transmission attempts are performed randomly, independently, distributively, based on a simple ACK/NACK feedback from the receiver. \cite{Abramson}. ALOHA has gained popularity due to its simple nature and because it does not require a centralized controller. 

The derivation of the stability region of random access systems for bursty sources is known to be a difficult problem. This is because each source transmits and interferes with the others only when its queue is non-empty. Such queues where the service process of one depends on the status of the others are said to be coupled or interacting. Thus, the individual departure rates of the queues cannot be computed separately without knowing the stationary distribution of the joint queue length process \cite{rao:stability}. There is a vast literature that considers stable throughput analysis in random access schemes \cite{SastryNOW, AEUnion, ALOHATIT2012, ASMTA2017}.

In the recent years, relaying and in general cooperative communications have received a great deal of attention. 
In \cite{PappasTWC2015, COMCOM2014, ACCESS2017}, the effect of a single relay in a wireless network is studied. These works consider half- and full-duplex cases and also the effect of relaying on delay. There are also works that consider multiple relays such as \cite{COMNET2015,VardheJCN2015, PloumidCOMCOM, dimitriou2016retrial, PloumidTVT, DIMITRIOU2017}. The relay selection is an important research direction in cases of multiple relays, and works that consider relay selection are 
\cite{ThemisTCOM2015, ThemisCL2015, ThemisSurvey, CL2017, JCN2017_Relay}. Papers that study the effect of energy harvesting on cooperative communications are \cite{Krikidis2012, Globalsip2013, JCN2012, JeonJSAC2015, JCN2016}.

\subsection{Contributions}
The main contribution in this paper is to introduce the notion of partial network-level cooperation by adding a flow controller for the traffic coming to the relay from the source. We prove that the system is always better than or at least equal to the system without the flow controller.
Specifically, we provide an exact characterization of the stability region of a network consisting of a source, a relay and a destination node as shown in Fig.~\ref{fig:model}.
We consider the collision channel with erasures and random access of the medium. The source and the relay node have external arrivals; furthermore, the relay is forwarding part of the source node's traffic to the destination. Unlike the work in~\cite{b:Rong3}, the relay node is equipped with a flow controller that regulates the internal arrivals from the source based on the conditions in the network to ensure the stability of the queues.
We characterize the stable throughput region under conditions of no cooperation at all, full cooperation, and probabilistic (opportunistic) cooperation. By probabilistic cooperation we mean that under certain conditions in the network, the relay may accept a packet from the source. The characterization of the stability regions is known to be challenging because the queues of the users are coupled (i.e., the service process of a queue depends on the status of the other queues). A tool that bypasses this difficulty is the stochastic dominance technique~\cite{rao:stability}.

\subsection{Organization}
In Section \ref{sec:model}, we describe the channel model, explain the cooperative model, and provide the definition of stability. In Section \ref{sec:main}, we provide the main theoretical results of this work. The proofs of these results are given in Sections \ref{sec:proofth0} and \ref{sec:proofth1}. In Section \ref{sec:results} we evaluate numerically the theoretical results. Finally, we conclude our paper in Section \ref{sec:conclusion}.

\section{System Model}\label{sec:model}

\begin{figure}[t]
\centering
\includegraphics[scale=0.3]{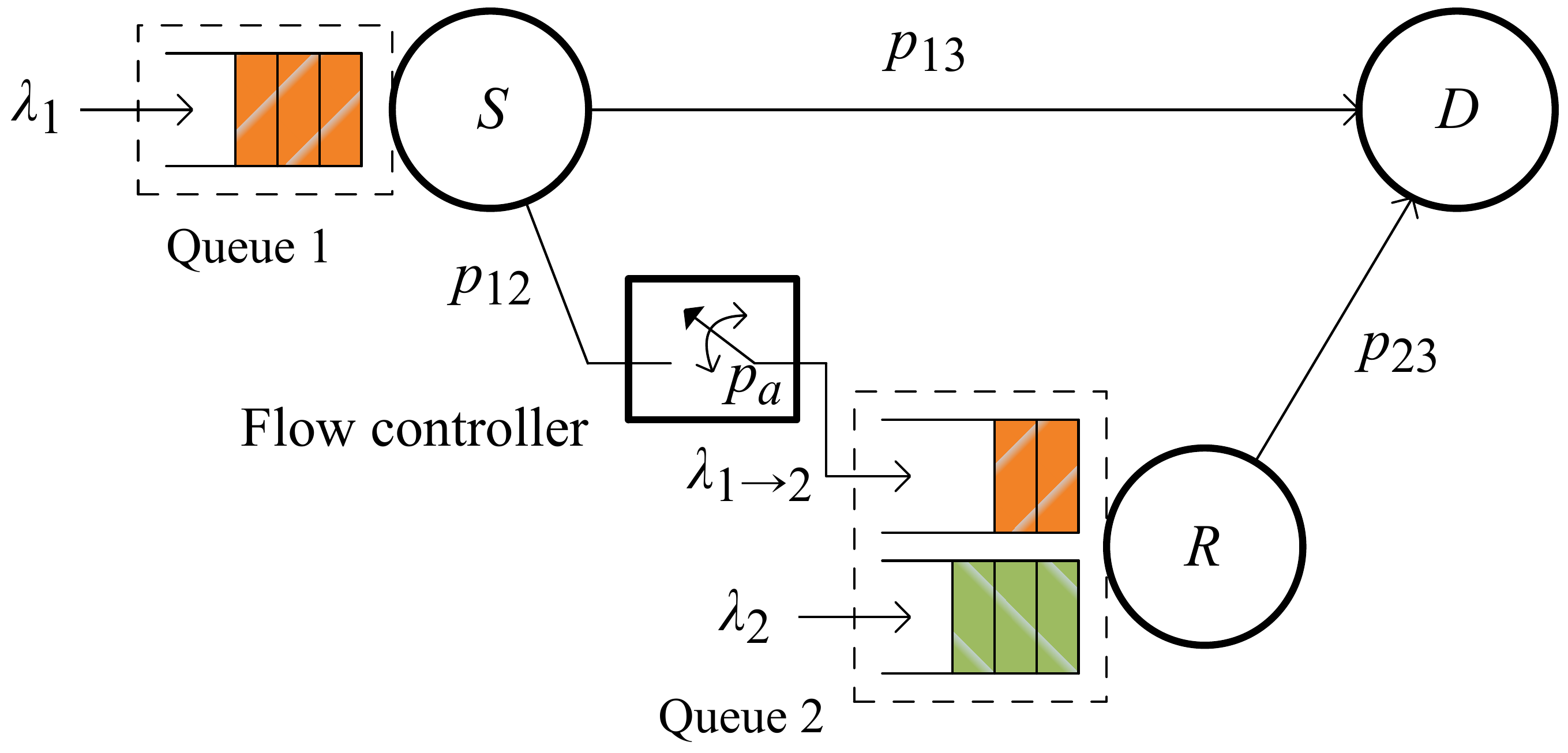}
\caption{A relay-aided wireless network. The relay is equipped with a flow controller for the incoming traffic from the source $S$.}
\label{fig:model}
\end{figure}

We consider a time-slotted system in which the nodes are randomly accessing a common receiver as shown in Fig.~\ref{fig:model}. We denote with $S$, $R$, and $D$ the source, the relay and the destination, respectively. Packet traffic originates from $S$ and $R$. Because of the wireless broadcast nature, $R$ may receive some of the packets transmitted from $S$ and then relay those packets to $D$. The packets from $S$ which failed to be received by $D$ but were successfully received by $R$ are relayed by $R$. As we impose the half-duplex constraint, $R$ can overhear $S$ only when it is idle. Each node has an infinite size buffer for storing incoming packets, and the transmission of each packet occupies one time slot. Node $R$ has separate queues for the exogenous arrivals and the endogenous arrivals that are relayed through $R$. But, we can let $R$ to maintain a single queue and merge all the arrivals into a single queue as the achievable stable throughput region is not affected~\cite{b:Rong3}. This is because the link quality between $R$ and $D$ is independent of which packet is selected for transmission.

The packet arrival processes at $S$ and $R$ are assumed to be Bernoulli with rates $\lambda_1$ and $\lambda_2$, respectively, and they are independent of each other. Node $R$ is equipped with a flow controller that regulates the rate of endogenous arrivals from $S$ by randomly accepting the incoming packets with probability $p_a$; that is, it controls the \textit{amount of cooperation} that it is willing to provide. In each time slot, nodes $S$ and $R$ attempt to transmit with probabilities $q_1$ and $q_2$, respectively, if their queues are not empty. Decisions on transmission are made independently among the nodes. We assumed collision channel with erasures in which, if both $S$ and $R$ transmit in the same time slot, a collision occurs and both transmissions fail. The probability that a packet transmitted by node $i$ is successfully decoded at node $j (\neq i)$ is denoted by $p_{ij}$ which is the probability that the signal-to-noise-ratio (SNR) over the specified link exceeds a certain threshold for the successful decoding. These erasure probabilities capture the effect of random fading at the physical layer. The probabilities $p_{13}$, $p_{23}$, and $p_{12}$ denote the success probabilities over the link $S-R$, $R-D$, and $S-R$, respectively. Node $R$ has a better channel to $D$ than $S$, that is $p_{23} > p_{13}$.

The cooperation is performed at the protocol level as follows. When $S$ transmits a packet, if $D$ decodes the packet successfully, it sends an ACK and the packet exits the network; if $D$ fails to decode the packet but $R$ does and the flow controller decides to relay the packet, then $R$ sends an ACK and takes over the responsibility of delivering the packet to $D$ by placing it in its queue. If neither $D$ nor $R$ decode (or if $R$ does not store the packet), the packet remains in $S$'s queue for retransmission. The ACKs are assumed to be error-free, instantaneous and broadcasted to all relevant nodes.

From the above we can compute the average service rates for the source and the relay. 
The average service rate for the source $S$, denoted by $\mu_{1}$, is
\begin{multline} \label{eqn:mu1}
\mu_{1} = \left\{ (1-q_{2}) \mathrm{Pr} (Q_{2} \neq 0) + \mathrm{Pr}(Q_{2} = 0) \right\} \\
\times q_{1} \left(p_{13}+(1-p_{13})p_{12}p_{a} \right).
\end{multline}
The average service rate for the relay $R$, denoted by $\mu_2$, is 
\begin{equation} \label{eqn:mu2}
\mu_{2}=q_{2} \left[1-q_1 \mathrm{Pr}( Q_1 \neq 0) \right]p_{23}.
\end{equation}

\subsection{Stability Criterion}

We use the following definition of queue stability~\cite{Szpankowski:stability}:

\begin{definition}
Denote by $Q_i^t$ the length of queue $i$ at the beginning of time slot $t$. The queue is said to be \emph{stable} if
$\lim_{t \rightarrow \infty} {Pr}[Q_i^t < {x}] = F(x)$ and $\lim_{ {x} \rightarrow \infty} F(x) = 1$.
\end{definition}

Although we do not make explicit use of this definition we use its corollary consequence which is Loynes' theorem~\cite{b:Loynes} that states that if the arrival and service processes of a queue are strictly jointly stationary and the average arrival rate is less than the average service rate, then the queue is stable. If the average arrival rate is greater than the average service rate, then the queue is unstable and the value of $Q_i^t$ approaches infinity almost surely. The stability region of the system is defined as the set of arrival rate vectors $\boldsymbol{\lambda}=(\lambda_1, \lambda_2)$ for which the queues in the system are stable.

\section{Main Results}\label{sec:main}

This section describes the stability region for the system presented in the previous section and depicted in Fig.~\ref{fig:model}.

The next theorem presents the stability region of the system for fixed probability of cooperation $p_a$.

\begin{theorem} \label{thm:th0}
The stability region of the partial cooperative network with fixed $p_a$ is described by the set of the arrival vectors $\boldsymbol{\lambda}=(\lambda_1, \lambda_2) \in \mathcal{R}(p_a)$, where $\mathcal{R}(p_a)=\mathcal{R}_{1}(p_a) \bigcup \mathcal{R}_{2}(p_a).$ The subregions $\mathcal{R}_{1}(p_a)$ and $\mathcal{R}_{1}(p_a)$ are given by \eqref{eqn:R1pa} and \eqref{eqn:R2pa} respectively. 
\end{theorem}

\begin{figure*}
\begin{align}
\mathcal{R}_1 (p_a)  =  \left\{ (\lambda_{1},\lambda_{2}) : \left[ 1+ \frac{p_{12}p_{a}(1-p_{13})  q_{1}}{(1- q_{1})p_{23}}  \right] \lambda_{1}+  \frac{ q_{1} \left[p_{12}p_{a}(1-p_{13}) +p_{13} \right]}{(1- q_{1})p_{23}} \lambda_{2}  <  q_{1} \left[ (1-p_{13})p_{12}p_{a}+p_{13} \right] , \right. \notag \\
\left. \lambda_{2}+\frac{(1-p_{13})p_{12}p_{a}} {p_{13}+(1-p_{13})p_{12}p_{a}} \lambda_{1} < q_{2} \left(1- q_{1} \right)p_{23} \right\} \label{eqn:R1pa} 
\end{align}
\begin{align}
\mathcal{R}_2 (p_a)  =  \left\{ (\lambda_{1},\lambda_{2}) : \lambda_2 + \frac{(1-q_2)(1-p_{13})p_{12}p_{a}+q_2 p_{23}}{(1-q_2)\left[p_{13}+(1-p_{13})p_{12}p_{a} \right]} \lambda_1 < q_2 p_{23}, \lambda_{1} < q_{1}(1-q_{2})\left[p_{13}+(1-p_{13})p_{12}p_{a} \right] \right\} \label{eqn:R2pa} 
\end{align}
\end{figure*}

\begin{proof}
The proof is given in Section \ref{sec:proofth0}.
\end{proof}

\begin{figure}[t]

\centering
 \subfigure[$\mathcal{R}_1(p_a)$ given by (\ref{eqn:R1pa}).]{
 \includegraphics[scale=0.4]{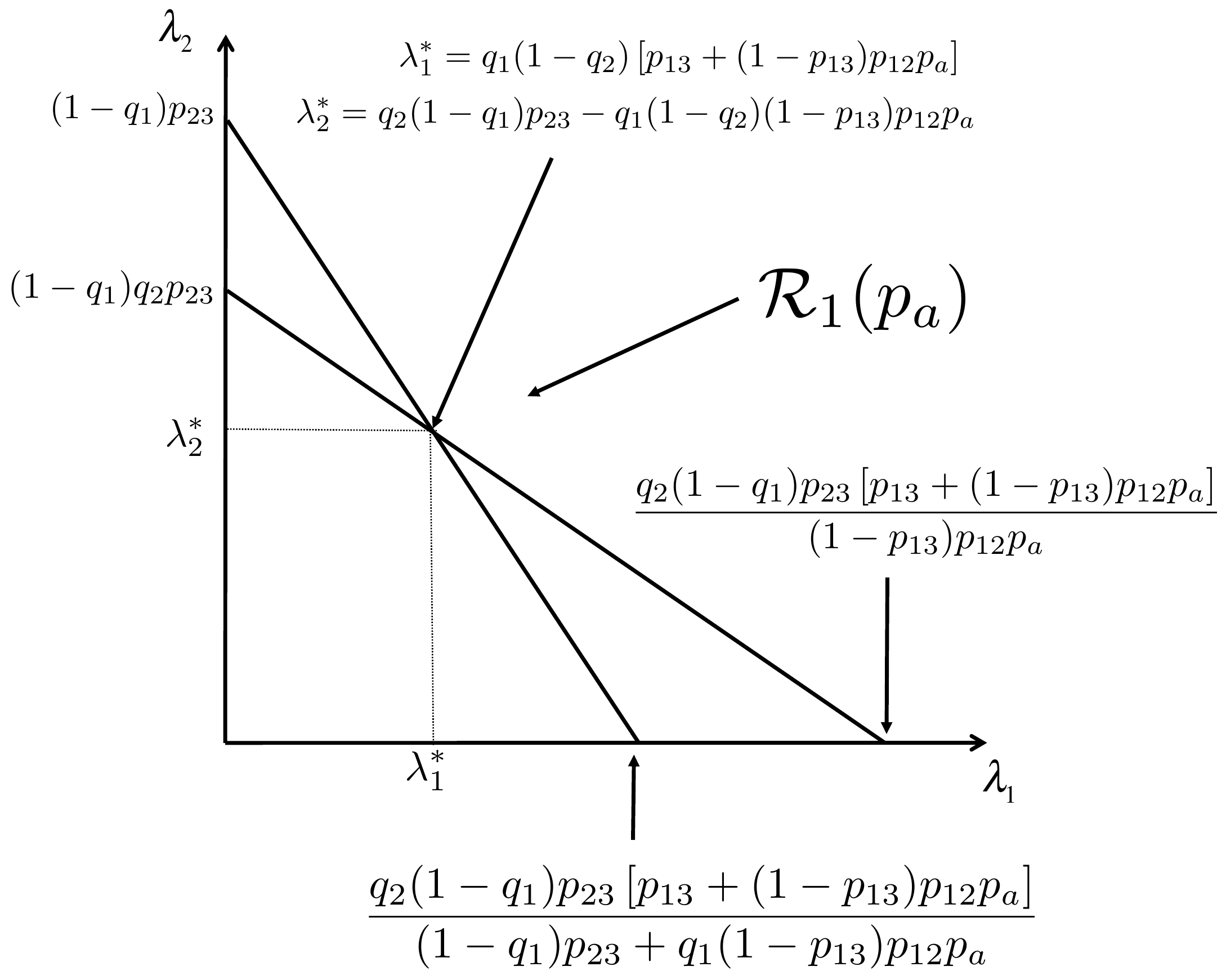}
 \label{fig:R1pa}
 }

  \subfigure[$\mathcal{R}_2(p_a)$ given by (\ref{eqn:R2pa}).]{
  \includegraphics[scale=0.4]{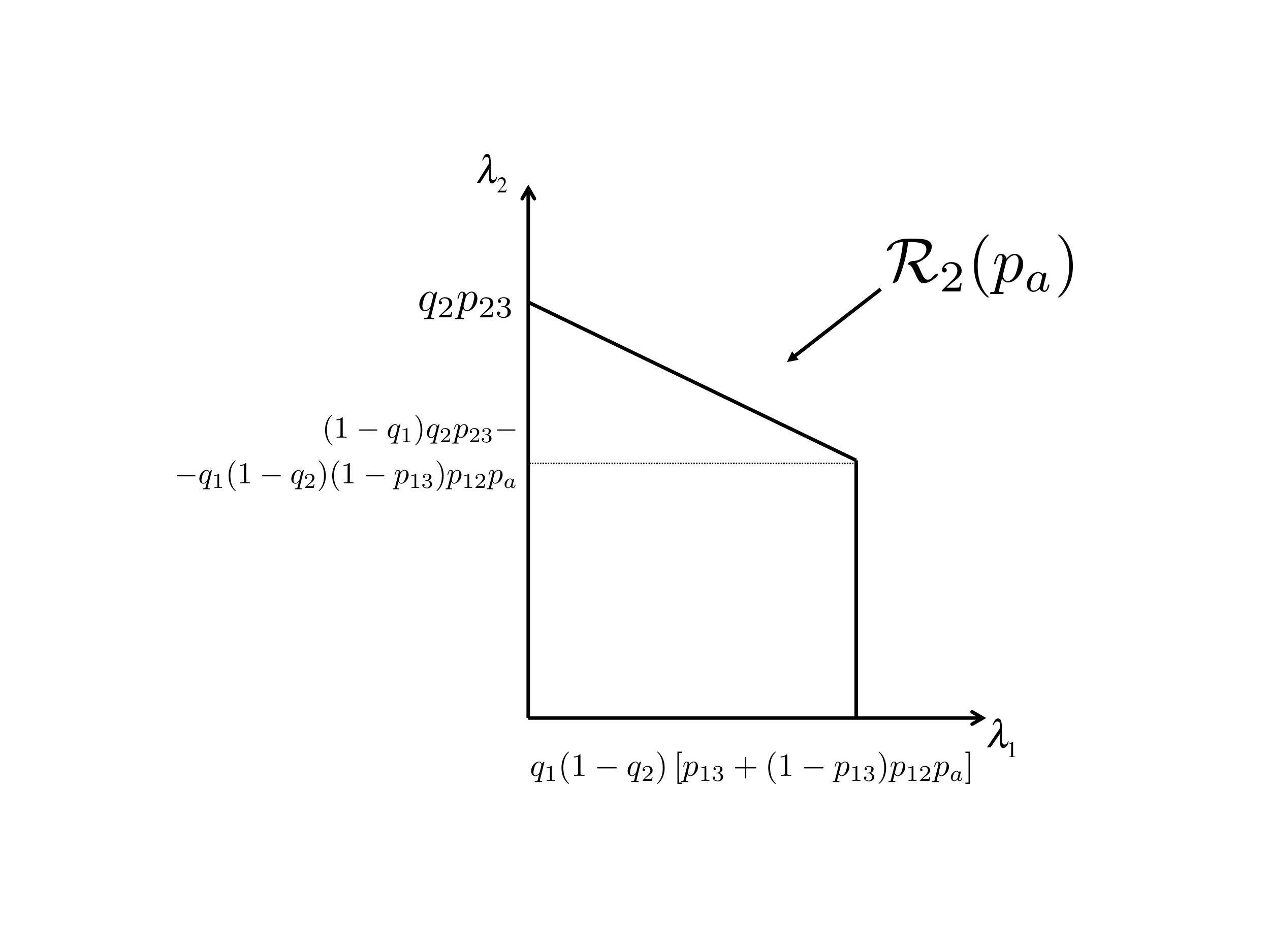}
  \label{fig:R2pa}
  }
  \caption{The stability region for fixed $p_a$, $\mathcal{R}(p_a) = \mathcal{R}_1 (p_a)\bigcup \mathcal{R}_2 (p_a)$, described in Theorem \ref{thm:th0}.}
\end{figure}

The subregions $\mathcal{R}_{1}(p_a)$ and $\mathcal{R}_{2}(p_a)$ are depicted in Fig. \ref{fig:R1pa} and Fig. \ref{fig:R2pa} respectively.

The previous theorem provided the stability region, $\mathcal{R}(p_a)$ for a given $p_a$, however, it will be of interest to find the closure of the stability region over all possible values of the amount of the cooperation. The closure of the stability region can be defined as

\begin{equation}
\mathcal{R} \triangleq \bigcup_{ p_a \in [0,1]} \mathcal{R}(p_a) = \bigcup_{ p_a \in [0,1]} \left( \mathcal{R}_{1}(p_a) \bigcup \mathcal{R}_{2}(p_a) \right).
\end{equation}

Thus, our objective is to find the optimum value of $p_a$ denoted by $p_a^{*}$ which maximizes the stability region. The next theorem provides the closure of the stability region over all the possible values of $p_a$.

\begin{theorem} \label{thm:th1}
The stability region of the opportunistic cooperative network depicted in Fig.~\ref{fig:model} is described by:
\begin{equation}
    \mathcal{R} = \mathcal{R}_1 \bigcup \mathcal{R}_2.
\end{equation}
\begin{itemize}
    \item The subregion $\mathcal{R}_1$ is described as follows:
    	\begin{itemize}
         	\item if $q_{1} < \frac{p_{23}}{ p_{13}+p_{23}}$, then $p_{a}^{*}=1$ and the region is given by (\ref{eqn:R11}).
        	\item if $q_{1} \geq \frac{p_{23}}{p_{13}+p_{23}}$, then $p_{a}^{*}=0$ and the region is given by (\ref{eqn:R12}).
        \end{itemize}
    \item The subregion $\mathcal{R}_2$ is described as follows:
        \begin{itemize}
            \item if $q_{2} \geq \frac{p_{13}}{p_{13}+p_{23}}$, then $p_{a}^{*}=1$ and the region is given by (\ref{eqn:R21}).
            \item if $q_{2} <\frac{p_{13}}{p_{13}+p_{23}}$, then the subregion $\mathcal{R}_2$ is $\mathcal{R}_2 = \mathcal{R}_2^{'} \bigcup \mathcal{R}_2^{''}$ where:
                \begin{itemize}
                    \item if $\lambda_{1} <  q_{1} (1-q_{2})p_{13}$, then $p_{a}^{*}=0$ and the region is given by (\ref{eqn:R221}).
                    \item if $\lambda_{1} \geq  q_{1} (1-q_{2})p_{13}$, then $p_{a}^{*}=\frac{\lambda_{1}- q_{1}(1-q_{2})p_{13}}{ q_{1} (1-q_{2})(1-p_{13})p_{12}}$ and the region is given by (\ref{eqn:R222}).
            \end{itemize}
    	\end{itemize}
\end{itemize}

\begin{figure*}
\begin{align}
\mathcal{R}_1 & =  \left\{ (\lambda_{1},\lambda_{2}) :  \frac{q_{1}p_{12}(1-p_{13})+(1- q_{1}) p_{23} }{ q_{1} [ p_{13} + (1  -  p_{13}) p_{12} ] } \lambda_{1} + {\lambda_{2}} <(1-q_{1})p_{23},  \frac{p_{12}(1-p_{13})}{p_{12}(1-p_{13})+p_{13}} \lambda_{1} +\lambda_{2} < q_{2} (1-q_{1}) p_{23} \right\} \label{eqn:R11} \\
\mathcal{R}_1 & =\left\{ (\lambda_{1},\lambda_{2}): \frac{\lambda_1}{q_1 p_{13}} + \frac{\lambda_2}{(1-q_1)p_{23}} < 1, \ \ \lambda_{2}<q_{2}(1-q_{1})p_{23} \right\} \label{eqn:R12} \\
\mathcal{R}_2 & =\left\{ (\lambda_{1},\lambda_{2}):\frac{(1-q_2)p_{12}(1-p_{13})+q_2 p_{23}}{(1-q_2)\left[p_{13}+(1-p_{13})p_{12} \right] }\lambda_1 + \lambda_2 < q_2 p_{23}, \ \ \lambda_1 < q_1 (1-q_2) \left[p_{13}+(1-p_{13})p_{12} \right] \right\} \label{eqn:R21} \\
\mathcal{R}_2^{'} & =\left\{ (\lambda_{1},\lambda_{2}): \frac{\lambda_1}{(1-q_{2})p_{13}} + \frac{\lambda_2}{q_{2}p_{23}}  < 1, \ \ \lambda_{1}< q_{1} (1-q_{2})p_{13} \right\} \label{eqn:R221}\\
                        \mathcal{R}_2^{''} & = \left\{ (\lambda_{1},\lambda_{2}):\lambda_1 +\lambda_2 < q_{1}(1-q_{2})p_{13}+q_{2} p_{23}(1-q_{1}),
                        \ \  q_{1} (1-q_{2})p_{13}\leq\lambda_{1}<q_{1}(1-q_{2})[p_{13}+(1-p_{13})p_{12}] \right\}  \label{eqn:R222}
\end{align}
\end{figure*}

\begin{proof}
The proof is given in Section \ref{sec:proofth1}.
\end{proof}
\end{theorem}

\begin{remark}
As seen in the Theorem \ref{thm:th1}, there are three possible optimal values of $p_a$. When $p_a^{*}$ equals to $0$ or $1$, the relay rejects or accepts all the incoming traffic from the source, respectively.
The more interesting case is when $q_{2} <\frac{p_{13}}{p_{13}+p_{23}}$ (the relay transmission probability is less than a threshold which is a function of the channel success probabilities) and at the same time the average arrival rate at the source is $\lambda_{1} \geq  q_{1} (1-q_{2})p_{13}$; in this case the optimum cooperation strategy is probabilistic routing by the relay. The incoming traffic from the source is relayed in part, meaning that the relay accepts a packet from the source with probability $p_a^{*}$, where $p_{a}^{*}=\frac{\lambda_{1}- q_{1}(1-q_{2})p_{13}}{ q_{1} (1-q_{2})(1-p_{13})p_{12}}$ ($0<p_a^{*}<1$).
The intuition behind this result is that when the relay is not attempting to transmit ``very often" and at the same time, the arrival rate at the source is greater than a certain value, then the relay is cooperating only partially. 
\end{remark}
\begin{remark}
When $q_{2} <\frac{p_{13}}{p_{13}+p_{23}}$ and $\lambda_{1} \geq  q_{1} (1-q_{2})p_{13}$, we proved that $p_{a}^{*}=\frac{\lambda_{1}- q_{1}(1-q_{2})p_{13}}{ q_{1} (1-q_{2})(1-p_{13})p_{12}}$, which can be rewritten as
\begin{equation}
p_{a}^{*}= \frac{1}{q_{1} (1-q_{2})} \left[\frac{\lambda_{1}}{(1-p_{13})p_{12}} - p_{13}\right].
\end{equation}
We can see that if $\lambda_{1}$ increases then $p_{a}^{*}$ increases too. If $q_{1}$ increases then $p_{a}^{*}$ decreases.
\end{remark}

\emph{From the above, it is clear that $p_{a}^{*}$ controls the amount of cooperation}.

\section{Proof of Theorem \ref{thm:th0}}
\label{sec:proofth0}

In this section we will prove the Theorem \ref{thm:th0} which describes the stability conditions of the system for given $p_a$.
The expressions for the average service rates seen by source $S$ and relay $R$ are given by \eqref{eqn:mu1} and \eqref{eqn:mu2} respectively.

Since the average service rate of each queue $\mu_{1}$ and $\mu_{2}$ depends on the queue size of the other queue, they cannot be computed directly. We bypass this difficulty by utilizing the idea of stochastic dominance~\cite{rao:stability}; that is, we first construct hypothetical dominant systems, in which one of the nodes transmits dummy packets even when its packet queue is empty. Since the queue sizes in the dominant system are, at all times, at least as large as those of the original system, the stability region of the dominant system inner-bounds that of the original system. It turns out, however, that the stability region obtained using this stochastic dominance technique coincides with that of the original system which will be discussed in detail later in this section. Thus, the stability regions for both the original and the dominant systems are the same.

\subsection{The first dominant system: source node transmits dummy packets}
In this sub-section we obtain the region $\mathcal{R}_1(p_a)$ of Theorem~\ref{thm:th0}.
We consider the first dominant system, in which node $S$ transmits dummy packets with probability $q_1$ whenever its queue is empty, while node $R$ behaves in the same way as in the original system. All other assumptions remain unaltered in the dominant system. Thus, the service rate at the relay node is given by:
\begin{equation} \label{eqn:mu2sd1}
\mu_{Q_{2}}=q_{2} \left(1-q_{1} \right)p_{23}.
\end{equation}
To derive the stability condition for the queue in the relay node, we need to calculate the total arrival rate. There are two independent arrival processes at the relay: the exogenous traffic with arrival rate $\lambda_2$ and the endogenous traffic from $S$. In the dominant system, when $R$ receives a dummy packet from $S$, it simply discards that packet. When the dominant system is stable, the queue at $S$ is stable, so the departure rate of the source packets (excluding the dummy ones) is equal to the arrival rate $\lambda_1$. Denote by $S_A$ the event that $S$ transmits a packet and the packet leaves the queue, then:
\begin{multline}
\mathrm{Pr}(S_A)=\left[(1-q_{2})\mathrm{Pr}(Q_{2} \neq 0) + \mathrm{Pr}(Q_{2}=0) \right] \\
\times \left[p_{13}+(1-p_{13})p_{12}p_{a} \right].
\end{multline}

Among the packets that depart from the queue of $S$, some will exit the network because they are decoded by the destination directly, and some will be relayed by $R$. Denote by $S_B$ the event that the transmitted packet from $S$ will be relayed from $R$, then:
\begin{equation}
\mathrm{Pr}(S_B)=\left[(1-q_{2})\mathrm{Pr}(Q_{2} \neq 0) + \mathrm{Pr}(Q_{2}=0) \right] (1-p_{13})p_{12}p_{a}.
\end{equation}
The conditional probability that a transmitted packet from $S$ (dummy packets excluded) arrives at $R$ given that the transmitted packet exits node $S$'s queue
is given by:
\begin{equation}
\mathrm{Pr}(S_B | S_A)=\frac{(1-p_{13})p_{12}p_{a}} {p_{13}+(1-p_{13})p_{12}p_{a}}.
\end{equation}
The total arrival rate at the relay node is:
\begin{equation}
\lambda_{Q_{2}}=\lambda_{2}+\frac{(1-p_{13})p_{12}p_{a}} {p_{13}+(1-p_{13})p_{12}p_{a}} \lambda_{1}.
\end{equation}
By Loyne's Theorem, the stability condition for queue $2$ at node $R$ is given by $\lambda_{Q_{2}} < \mu_{Q_{2}}$ and, thus:
\begin{equation}\label{eqn:R1pa2}
\lambda_{2}+\frac{(1-p_{13})p_{12}p_{a}} {p_{13}+(1-p_{13})p_{12}p_{a}} \lambda_{1} < q_{2} \left(1- q_{1} \right)p_{23}.
\end{equation}
The probability that the queue is not empty can be computed by Little's theorem and is given by:
\begin{equation} \label{eqn:probQ2}
\mathrm{Pr} (Q_{2} \neq 0) = \frac{\lambda_{Q_{2}}}{\mu_{Q_{2}}}=\frac{\lambda_{2}+\frac{(1-p_{13})p_{12}p_{a}} {p_{13}+(1-p_{13})p_{12}p_{a}} \lambda_{1}}{q_{2} \left(1- q_{1} \right)p_{23}}.
\end{equation}
Thus, after substituting (\ref{eqn:probQ2}) into (\ref{eqn:mu1}), the average service rate seen by $S$ is
\begin{multline}
\mu_{1} = \frac{ q_{1}}{(1- q_{1})p_{23}} \left\{ \left[p_{12} p_{a} (1-p_{13}) + p_{13} \right] (1-q_{1})p_{23} \right.\\
\left. {} -p_{12}(1-p_{13})p_{a}\lambda_{1}-\left[p_{12}(1-p_{13})p_{a} +p_{13} \right]\lambda_{2}  \right\}.
\end{multline}
The stability condition for queue $1$ at the source node is $\lambda_{1} < \mu_{1}$, and after some algebra, we obtain:
\begin{multline} \label{eqn:R1pa1}
\left[ 1+ \frac{p_{12}p_{a}(1-p_{13})  q_{1}}{(1- q_{1})p_{23}}  \right] \lambda_{1}+  \frac{ q_{1} \left[p_{12}p_{a}(1-p_{13}) +p_{13} \right]}{(1- q_{1})p_{23}} \lambda_{2} \\ <  q_{1} \left[ (1-p_{13})p_{12}p_{a}+p_{13} \right].
\end{multline}

The derived stability conditions, $\mathcal{R}_1(p_a)$ , obtained from the first dominant system are summarized in \eqref{eqn:R1pa}.

An important observation made in \cite{rao:stability} is that the stability conditions obtained by using the stochastic dominance technique are not merely sufficient conditions for the stability of the original system but are sufficient and necessary conditions. The \emph{indistinguishability} argument applies to our problem as well. Based on the construction of the dominant system, it is easy to see that the queues of the dominant system are always larger in size than those of the original system, provided they are both initialized to the same value. Therefore, given $\lambda_{2}<\mu_{2}$, if for some $\lambda_{1}$, the queue at $S$ is stable in the dominant system, then the corresponding queue in the original system must be stable; conversely, if for some $\lambda_{1}$ in the dominant system, the queue at node $S$ saturates, then it will not transmit dummy packets, and as long as $S$ has a packet to transmit, the behavior of the dominant system is identical to that of the original system because the dummy packet transmissions are increasingly rare as we approach the stability boundary. Therefore, we can conclude that the original system and the dominant system are indistinguishable at the boundary points.

\subsection{The second dominant system: relay node transmits dummy packets}

In this sub-section we obtain the region $\mathcal{R}_2(p_a)$ of Theorem~\ref{thm:th0}.
We consider the second dominant system, in which node $R$ transmits dummy packets with probability $q_2$ whenever its queue is empty, while node $S$ behaves in the same way as in the original system. All the other assumptions remain unaltered in the dominant system. The service rate for the source node is
\begin{equation}
\mu_{1}= q_{1}(1-q_{2})\left[p_{13}+(1-p_{13})p_{12}p_{a} \right].
\end{equation}
Thus, queue $1$ is stable if
\begin{equation} \label{eqn:lambda1sd2}
\lambda_{1} < q_{1}(1-q_{2})\left[p_{13}+(1-p_{13})p_{12}p_{a} \right].
\end{equation}
The probability that the queue is not empty is:
\begin{equation}
\mathrm{Pr}(Q_1 \neq 0) = \frac{\lambda_1}{\mu_{1}}=\frac{\lambda_1}{q_{1}(1-q_{2})\left[p_{13}+(1-p_{13})p_{12}p_{a} \right]}.
\end{equation}
The total arrival rate at the relay node is given by:
\begin{equation}
\lambda_{Q_{2}}=\lambda_{2}+\mathrm{Pr}(S_B|S_A) \lambda_{1},
\end{equation}
where $S_A$ and $S_B$ are defined in the previous sub-section. Note that $\mathrm{Pr}(S_A)=(1-q_{2}) \left(p_{13}+(1-p_{13})p_{12}p_{a} \right)$, $\mathrm{Pr}(S_B)=(1-q_{2})(1-p_{13})p_{12}p_{a}$ and, thus, we have $\mathrm{Pr}(S_B|S_A)=\frac{(1-p_{13})p_{12}p_{a}} {p_{13}+(1-p_{13})p_{12}p_{a}}$.
From the above it follows that the total arrival rate at the relay node is:
\begin{equation}
\lambda_{Q_{2}}=\lambda_{2}+\frac{(1-p_{13})p_{12}p_{a}} {p_{13}+(1-p_{13})p_{12}p_{a}} \lambda_{1}.
\end{equation}
The service rate for the relay node is:
\begin{align}
\mu_{Q_2}=q_{2} \left[1-q_1 \mathrm{Pr}( Q_1 \neq 0) \right]p_{23}.
\end{align}
Thus, from Loyne's stability criterion, it follows that the queue is stable if $\lambda_{Q_{2}}<\mu_{Q_2}$ and, thus:
\begin{equation}
\lambda_{2}+\frac{(1-p_{13})p_{12}p_{a}} {p_{13}+(1-p_{13})p_{12}p_{a}} \lambda_{1} < q_{2} \left[1-q_1 \mathrm{Pr}(Q_1 \neq 0) \right]p_{23}.
\end{equation}
After some algebra, we obtain:
\begin{equation}
\lambda_2 + \frac{(1-q_2)(1-p_{13})p_{12}p_{a}+q_2 p_{23}}{(1-q_2)\left[p_{13}+(1-p_{13})p_{12}p_{a} \right]} \lambda_1 < q_2 p_{23}.
\end{equation}
The derived stability conditions, $\mathcal{R}_2(p_a)$, obtained from the second dominant system are summarized in \eqref{eqn:R2pa}.

The indistinguishability argument at saturation holds here as well.
This concludes the proof of Theorem \ref{thm:th0}.

\section{Proof of Theorem \ref{thm:th1}}
\label{sec:proofth1}
In this section we will give the proof of Theorem \ref{sec:proofth1} which provides the closure of the stability region over all the possible values of $p_a$. We will utilize the results of the previous section. In order to obtain the closure we have to solve two optimization problems.
\subsection{Derivation of $\mathcal{R}_1$}
Now we will find the value of $p_{a}$ that maximizes $\lambda_1$. After replacing $\lambda_1$ with $y$ and $\lambda_2$ with $x$ into \eqref{eqn:R1pa1} and \eqref{eqn:R1pa2} from the fist dominant system in Section \ref{sec:proofth0}, we have the following optimization problem:
\begin{multline} \label{eqn:ysd1}
\text{[$P_1$]       }   \max_{p_a} \text{   } y=\frac{- q_1 \left[p_{13}+(1-p_{13})p_{12}p_{a} \right]}{ q_{1}p_{12}(1-p_{13})p_{a}+(1- q_{1})p_{23}}x \\
+\frac{ q_1 \left[p_{13}+(1-p_{13})p_{12}p_{a} \right](1- q_1)p_{23}}{ q_{1}p_{12}(1-p_{13})p_{a}+(1- q_{1})p_{23}},
\end{multline}
subject to
\begin{equation} \label{eqn:xsd1}
0 \leq x \leq q_{2}(1- q_{1})p_{23} - \frac{p_{12}p_{a}(1-p_{13})}{p_{13}+(1-p_{13})p_{12}p_{a}}y,
\end{equation}
\begin{eqnarray}
& p_a \in [0,1], \\
& (x, y) \in [0,1]^2.
\end{eqnarray}
After differentiating $y$ with respect to $p_{a}$, we have
\begin{equation}
\frac{dy}{dp_{a}}= \left( \frac{A}{B} \right)^{'}=\frac{A^{'}B-AB^{'}}{B^2},
\end{equation}
where $B= q_{1}p_{12}(1-p_{13})p_{a}+(1- q_{1})p_{23}$ and
\begin{multline}
A^{'}B-AB^{'}= (1-p_{13})p_{12}q_{1}(x-p_{23}+ p_{23}q_1) \\
\times ( p_{13}q_{1}-p_{23}+ q_{1} p_{23}).
\end{multline}
From (\ref{eqn:mu2sd1}), it is obvious that $x-p_{23}+ p_{23} q_{1} < 0$.
If $ p_{13}q_{1}-p_{23}+ p_{23} q_{1} <0$, then we have that $q_{1} < \frac{p_{23}}{p_{13}+p_{23}}$. Then, $\frac{dy}{dp_{a}}>0$ and
$y$ is an increasing function of $p_{a}$ and, thus $p_{a}^{*}=1$. Then, (\ref{eqn:xsd1}) becomes
\begin{equation}
0 \leq x \leq q_{2}(1- q_{1})p_{23} - \frac{p_{12}(1-p_{13})}{p_{13}+(1-p_{13})p_{12}}y,
\end{equation}
and (\ref{eqn:ysd1}) becomes
\begin{multline}
y=\frac{-q_1 \left[p_{13}+(1-p_{13})p_{12} \right]}{ q_{1}p_{12}(1-p_{13})+(1- q_{1})p_{23}}x \\
+\frac{ q_1 \left[p_{13}+(1-p_{13})p_{12} \right](1- q_1)p_{23}}{ q_{1}p_{12}(1-p_{13})+(1- q_{1})p_{23}}.
\end{multline}
The stability region for this case is given by (\ref{eqn:R11}).
If $q_{1}>\frac{p_{23}}{p_{13}+p_{23}}$, it follows that $\frac{dy}{dp_{a}}<0$ and, thus, $y$
is a decreasing function of $p_{a}$ and $p_{a}^{*}=0$.
Then (\ref{eqn:xsd1}) becomes
\begin{equation}
0 \leq x \leq q_{2}(1- q_{1})p_{23},
\end{equation}
and (\ref{eqn:ysd1}) becomes
\begin{equation}
y+\frac{ q_{1}p_{13}}{(1- q_1)p_{23}}x = q_{1}p_{13}.
\end{equation}
The stability region is given by (\ref{eqn:R12}).

\subsection{Derivation of $\mathcal{R}_2$}
Next we find the value of $p_{a}$ that maximizes $\lambda_2$. After replacing $\lambda_1$ with $x$ and $\lambda_2$ with $y$ from the results obtained from the second dominant system in Section \ref{sec:proofth0}, we have the following optimization problem:
\begin{equation}
\text{[$P_2$]       }\max_{p_a} \text{   }y=q_2 p_{23} - \frac{(1-q_{2})p_{12}(1-p_{13})p_{a}+q_{2}p_{23}}{(1-q_{2})\left[p_{13}+(1-p_{13})p_{12}p_{a} \right]}x.
\end{equation}
subject to
\begin{equation}
x < q_{1}(1-q_{2})\left[p_{13}+(1-p_{13})p_{12}p_{a} \right],
\end{equation}
\begin{eqnarray}
& p_a \in [0,1], \\
& (x, y) \in [0,1]^2.
\end{eqnarray}
After differentiating $y$ with respect to $p_{a}$ we have
\begin{equation}
\frac{dy}{dp_{a}}= \left( \frac{A}{B} \right)^{'}=\frac{A^{'}B-AB^{'}}{B^2},
\end{equation}
where:
\begin{equation}
A^{'}B-AB^{'}=x p_{12}(1-p_{13})(1-q_{2})(p_{13}q_{2}-p_{13}+q_{2}p_{23}).
\end{equation}

If $p_{13}q_{2}-p_{13}+q_{2}p_{23}>0$, it follows that $\frac{p_{13}}{p_{13}+p_{23}}<q_{2}<1$ and $y$ increases; thus $p_{a}^{*}=1$ and, therefore:
\begin{equation}
x< q_{1} (1-q_{2}) \left[p_{13}+(1-p_{13})p_{12} \right],
\end{equation}
and
\begin{equation}
y+\frac{(1-q_{2})p_{12}(1-p_{13})+q_{2}p_{23}}{(1-q_{2})\left[p_{13}+(1-p_{13})p_{12} \right]}x=q_2 p_{23}.
\end{equation}
Then, the stability region is given by (\ref{eqn:R21}). If $q_{2} < \frac{p_{13}}{p_{13}+p_{23}}$, it follows that $y$ decreases and thus $p_{a}^{*}=0$, hence:
\begin{equation}
x< q_{1} (1-q_{2})p_{13},
\end{equation}
and
\begin{equation}
y+\frac{q_{2}p_{23}}{(1-q_{2})p_{13}}x=q_{2}p_{23}.
\end{equation}
The stability region is given by (\ref{eqn:R221}).

If $q_{1} (1-q_{2})p_{13} \leq x \leq  q_{1} (1-q_{2}) \left[p_{13}+(1-p_{13})p_{12}p_{a} \right]$, it follows from (\ref{eqn:lambda1sd2}) that
$p_{a} \geq \frac{x- q_{1} (1-q_{2})p_{13}}{ q_{1} (1-q_{2})(1-p_{13})p_{12}}$ and, thus we obtain that:
\begin{equation}
p_{a}^{*}=\frac{x- q_{1} (1-q_{2})p_{13}}{ q_{1} (1-q_{2})(1-p_{13})p_{12}},
\end{equation}
and
\begin{equation}
x+y=p_{23}q_{2}+ q_{1} (1-q_{2}) p_{13} - q_{1} q_{2} p_{23}.
\end{equation}
Finally, since $0 \leq p_{a} \leq 1$ we have that
\begin{equation}
x \leq  q_{1} (1-q_{2}) \left[p_{13}+(1-p_{13})p_{12} \right],
\end{equation}
and the stability region is given by (\ref{eqn:R222}).

This concludes the proof of Theorem \ref{thm:th1}.

\section{Numerical Results} \label{sec:results}

\begin{figure}[t]
\centering
\includegraphics[scale=0.5]{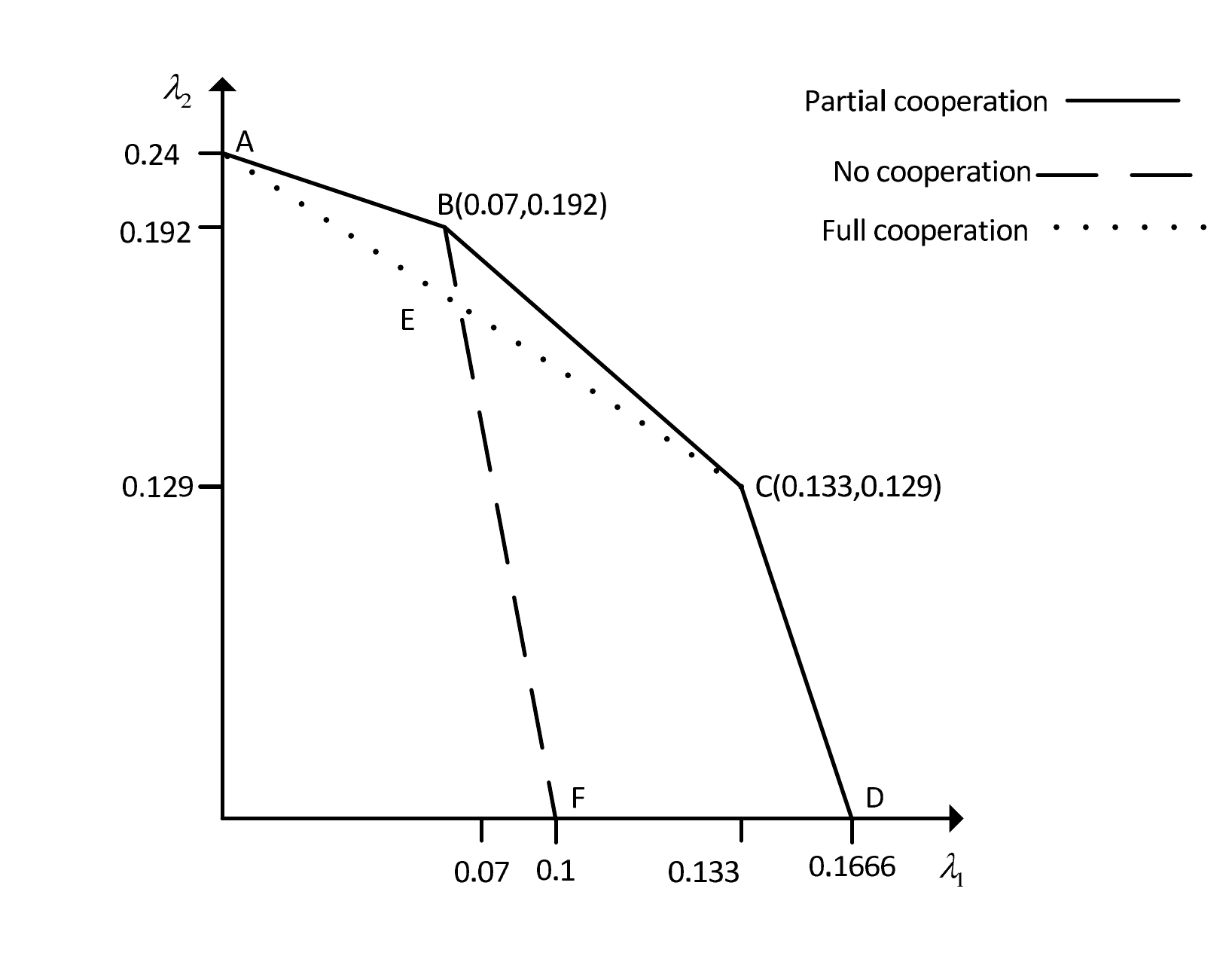}
\caption{Illustration of the stability region ($q_1=0.2$, $q_2=0.3$, $p_{13}=0.5$, $p_{12}=0.9$ and $p_{23}=0.8$).}
\label{fig:region}
\end{figure}

In this section, we obtain the stability region for the three cases of no-cooperation, full cooperation and partial cooperation and we compare them in a numerical illustration where $p_a < 1$. We let $q_1=0.2$, $q_2=0.3$, $p_{13}=0.5$, $p_{12}=0.9$, and $p_{23}=0.8$. In Fig.~\ref{fig:region}, we show the stability regions for the three cases. The region of partial cooperation contains the regions of the other cases. The boundaries of the stability region for the partial cooperation scheme are described by the line segments $ABCD$, and contains the region of non-cooperation ($ABF$) and the full cooperation ($ACD$). The triangular area $BEC$ in Fig.~\ref{fig:region} is achieved only by the partial cooperation scheme, showing that this scheme is superior compared to the other schemes. Note that in order to obtain the stability region for a fixed value of $p_a$ one can use the results from Theorem \ref{thm:th0} as depicted in Figs. \ref{fig:R1pa} and \ref{fig:R2pa}.

The line segment $AB$ belongs to the stability region of both no-cooperation and partial cooperation schemes. It is the boundary when $\lambda_1 \leq 0.07$ (which is the average arrival rate at the source) and, corresponds to the scheme of no-cooperation or when $p_a^{*}=0$. The line segment $CD$ is the boundary for the stability region for full cooperation and partial cooperation with $p_{a}^{*}=1$ when $0.133 \leq \lambda_1 < 0.1666$. The most interesting case is the $BC$ segment. This boundary is achieved only by the partial cooperation scheme. The value of $p_a^{*}$ that achieves the boundary is $p_a^*=\frac{\lambda_1 - 0.07}{0.063}$, as $0.07 < \lambda_1 < 0.133$. In this case the relay, through the flow controller, regulates the endogenous traffic from the source by randomly accepting (with $p_a^{*}$) the packets from the source. Note that as $\lambda_1$ increases (in the interval $(0.07,0.133)$) so does $p_a^{*}$. The values of $p_{a}^{*}$ that achieve the boundaries of the regions are given in Table~\ref{table}.

\begin{table}
\caption{The values of $p_{a}^*$}
\begin{center}\label{table}
    \begin{tabular}{ | l | l |}
    \hline
    Line & $p_a^{*}$ \\ \hline
    $AB$ & $0$ \\ \hline
    $BF$ & $0$  \\ \hline
    $AC$ & $1$  \\ \hline
    $CD$ & $1$  \\ \hline
    $BC$ & $\frac{\lambda_1 - 0.07}{0.063}$  \\ \hline
    \end{tabular}
\end{center}
\end{table}

The intuition behind these results, is that when the traffic level at the source is relatively low, the optimal scheme for the relay is not to cooperate at all. When the traffic level at the source is high, the best scheme is to fully cooperate, Finally, when the source has an intermediate level of traffic, the optimal scheme is to partially offer relay services.

\section{Conclusion} \label{sec:conclusion}
In this work, we introduced the notion of partial network-level cooperation by assuming a flow controller for the endogenous traffic to the relay from the source node of the network in Fig.~\ref{fig:model}. We provided an exact characterization of the stability region for this network. We proved that the system with the flow controller is always better than or at least equal to the system without the flow controller. The flow controller regulates the degree of cooperation offered by the relay.
Future directions of this work, are to consider multi-packet reception instead of collisions, and also to consider applying the notion of partial cooperation in larger topologies.

\bibliographystyle{IEEEtran}
\bibliography{bibliography}
\end{document}